# Mental Perception of Quality:

# Green Marketing as a Catalyst for Brand Quality Enhancement

**Saleh Ghobbe**

Department of Entrepreneurship, Faculty of Entrepreneurship, University of Tehran

**Mahdi Nohekhan**

Executive Management Group, Islamic Azad University, Shahrood Branch,

**Abstract**

The environmental conservation issue has led consumers to rethink the products they purchase. Nowadays, many consumers are willing to pay more for products that genuinely adhere to environmental standards to support the environment. Consequently, concepts like green marketing have gradually infiltrated marketing literature, making environmental considerations one of the most important activities for companies. Accordingly, this research investigates the impacts of green marketing strategy on perceived brand quality (case study: food exporting companies). The study population comprises 345 employees and managers from companies such as Kalleh, Solico, Pemina, Sorbon, Mac, Pol, and Casel. Using Cochran's formula, a sample of 182 individuals was randomly selected. This research is practical; the required data were collected through surveys and questionnaires. The findings indicate that (1) green marketing strategy has a significant positive effect on perceived brand quality, (2) green products have a significant positive effect on perceived brand quality, (3) green promotion has a significant positive effect on perceived brand quality, (4) green distribution has a significant positive effect on perceived brand quality, and (5) green pricing has a significant positive effect on perceived brand quality.

## Introduction

In the not-so-distant past, the industrial world experienced rapid growth in meeting consumer needs. Quick response to the consumer market, product, and service diversification, creating customer communication systems, and thousands of other achievements led organizations to perceive themselves as on the right path to excellence and success. However, environmental changes such as environmental degradation, air pollution, ozone layer depletion, and gradual global warming were unfortunate consequences of this industrial approach, gradually causing a massive wave of concerns and doubts and confronting industrial leaders with the harsh reality of production, supply, and consumption at the cost of Earth's destruction. The necessity of addressing environmental issues becomes clearer when we realize that sustainable development is impossible without considering the environment (Nawaser et al., 2023). In 1995, Polonsky defined green or environmental marketing as all activities designed to facilitate exchanges to satisfy human needs and desires in such a way that the satisfaction of these needs and desires involves minimal detrimental and destructive environmental impacts. Therefore, in light of the challenges facing most companies, green marketing strategies are implemented by developing and creating various green marketing programs (Karsten, Iris, 2014; Hessari et al., 2023; Ghasemi Kooktapeh et al., 2023). Green marketing is a broader concept that can be applied to consumer goods, industrial products, or even services.

As previously mentioned, unfortunately, most people believe that green marketing refers exclusively to promoting or advertising products with environmental features. Terms like "phosphate-free," "recyclable," and "ozone-friendly" are often associated with green marketing by consumers. However, these terms are just indicators of green marketing. Green marketing is a broader concept that can be applied to consumer goods, industrial products, and even services (Polonsky). Green marketing refers to developing and improving pricing, promoting, and distributing products that do not harm the environment (Pride & Ferrell, 1995).

The growth of green marketing and the demand for green products "may be the greatest opportunity for investment and the invention of new products in today's saturated industrial world." Today, organizations worldwide believe they have an ethical commitment and social responsibility. For example, costs associated with waste disposal and efforts to reduce them encourage companies to change their behavior. Companies with environmental competitors are under pressure to change their marketing activities. This emphasis on ethical commitment and social responsibility has made government organizations more accountable to the people and the environment (Mohajan, 2012).

Today, the determining environmental factor describes the world's public interest in societal welfare as follows:

- Interest in the purity of air, soil, and its resources.
- Protection and conservation of nature by using natural resources with an emphasis on marketing (recycling).
- Conservation in the use of non-renewable materials.
- Advancement in using waste materials in the production of new products.
- Spread awareness of the environment and have a healthy life (Alipour et al., 2011).

## Green Pricing

Green pricing programs focus on financial pricing methods that account for environmental and economic costs in production and marketing, providing value to consumers and a suitable profit for traders (Glen, Sandink, 2014). Price is a key and important factor in the green marketing mix. Most consumers are only willing to pay a higher price if they perceive added value in the product. This value may be in improved performance, efficiency, design, visual appeal, or taste, or even due to other features of the green product, such as longer life and harmlessness. However, it is essential to note that green pricing must be logical and competitive (Esmailpour et al., 2010).

**Green Product**

Ecological objectives in product design lead to reduced resource consumption and pollution and increased conservation of scarce resources. A green product helps preserve and improve the natural environment by conserving energy or resources and reducing or eliminating the use of toxic materials, pollution, and waste. In essence, a green product causes less environmental harm and is achieved through repair, renovation, reproduction, reuse, recycling, and reduction methods. Generally, the green and sustainable features of products and services can be summarized as:

- Designed to satisfy real human needs.
- Harmless to human health.
- Green throughout its lifecycle (Esmailpour et al., 2010).

**Green Distribution**

Green distribution programs include monitoring and enhancing environmental performance in the company's supply chain. Green promotion programs reflect communications designed to inform stakeholders about the company's efforts, commitment, and successes in environmental conservation.

Green distribution encompasses two dimensions: internal and external. The internal dimension refers to the company's internal environment, which should not only comply with environmental issues in internal processes and match the interior design with the product but also create a relaxing atmosphere for managers and employees, attracting customers with good employee interactions and a pleasant environment. The external dimension refers to distribution locations that cause the least environmental harm.

Reverse logistics topics (based on systems to aid material recycling) can also be considered under green distribution. With reverse logistics, the following aspects are important for companies:

- Identification: Tracking goods through the reverse logistics process.
- Recycling: Collecting goods for reprocessing.
- Reviewing: Testing materials to determine if they can meet reprocessing standards or need to be separated or disposed of.
- Reproducing: Reproducing the product according to its original standards or separating suitable parts for reuse.
- Eliminating: Disposing of materials that are not reproducible and selling reprocessed goods to existing or new customers.
- Re-engineering: Assessing existing goods for better design (Esmailpour et al., 2010).

**Green Promotion Programs**

Green promotion programs reflect communications designed to inform stakeholders about a company's efforts, commitment, and successes in environmental conservation. Such efforts may include advertising environmental appeals and requests, publicizing environmental efforts, and integrating environmental requests with product packaging.

Green promotion means conveying actual environmental information to consumers associated with the company's activities. It also represents the company's commitment to conserving natural resources to attract the target market.

A company should develop an integrated communications approach encompassing both "company-specific" and "product-specific" aspects of environmental issues and social responsibility. The strategies and slogans used by the company should be based on accurate research and information. Information should be coherently and consistently communicated to customers and other stakeholders, and companies should be cautious and vigilant about any communication that might seem exaggerated. Consumers, pressure groups, and the media are important audiences. If the slogans are incorrect and unchecked, the promotion will have adverse results. Ambitious and vague advertisements or slogans that do not resonate with the product and company are at risk of damaging customer perceptions (Esmailpour et al., 2010).

**Perceived Quality**

Perceived quality is considered the most critical competitive advantage for most organizations, whether in production or services. Therefore, organizations strive to improve profitability, reduce costs logically, maintain and increase market share, enhance customer satisfaction, and continually find innovative ways to improve the quality of their products and services. Perceived quality can be defined as the customer's perception of a product or service's overall quality or superiority compared to other options. Ultimately, perceived quality is a general and intangible feeling about a brand. However, perceived quality is usually based on key dimensions, including product characteristics such as reliability and performance, which are inherently linked to the brand. Identifying and measuring these main dimensions will be beneficial for understanding perceived quality, but it should not be forgotten that perceived quality is, after all, a general perception (Karbasi et al., 2011; Daneshmandi et al., 2023).

Zeithaml (1998) defined perceived quality as the customer's perception of the superior quality of a product or service compared to competitors, which does not include the technical aspect. He also indicates that perceived quality is part of a brand's unique value. Thus, high perceived quality leads consumers to choose one brand over competing brands. Therefore, the brand's unique value increases as perceived quality increases. Consumer perceived quality is associated with the evaluation of information and loyalty to a brand and significantly impacts the consumer's purchasing stage. Gill (2007) states that children in a family experience the quality of brands that their parents consume or recommend to them, and their perception of quality is due to their parents' knowledge of brands, which is more than their own. This experience has a long-lasting effect on them. Perceived quality may also guide an individual in purchasing a product or being loyal to a brand, which is something in the realm of intelligence and personal satisfaction (Bai et al., 2023). However, this hypothesis was not proven in his research. Aaker defines perceived quality as the customer's perception of a product or service's overall quality or superiority compared to other options. In his model, he states that perceived quality can affect the unique value of a brand in 5 ways: (1) a reason to buy the brand, (2) differentiation/positioning, (3) willingness to pay a premium price, (4) attracting the interest of distribution channel members to use a product with higher perceived quality, (5) brand development (Roosta and Madani, 2010).

**Perceived Quality Definition**

Service quality is defined as the comparison customers make between their expectations and their perception of the received services (Lai & Kwang, 2012).

Four definitions of service quality have been proposed:

- The alignment of the service with the desired characteristics of customers.
- The extent to which the service can satisfy the customer.
- A fair balance between the price and value of the service.
- The suitability of services for use (Roosta and Madani, 2010).

Alongside this concept, perceived quality is introduced based on the individual and mental perception of the visitor from the provided services coming from a mental capability that is influenced by the level of mental health (Bai & Vahedian, 2023). Perceived quality is not the actual quality of the product but the customer's mental evaluation of the product (Imani et al., 2010).

Some experts believe that perceived quality is the degree of alignment between perceived performance and customer expectations. Other researchers consider perceived quality as a result of satisfaction. Perceived value can be positively influenced by perceived quality. However, there is only sometimes a positive correlation between a customer's perception of quality and their perception of value. Customers may perceive high value from low-quality products or services due to low prices. Perceived quality is often defined as the consumer's judgment of the overall value of a product or service according to their objectives. It can also be seen as a customer's belief or attitude regarding the overall advantage of a product or service (Imani et al., 2010; Hessari, 2023).

Perceived quality is "the consumer's perception of the overall quality or superiority of a product or service compared to other options." In other words, perceived quality is a general and intangible (non-tangible) feeling about a brand. This factor goes beyond the actual quality of the brand. Even if a brand has desirable quality, it may not have created a favorable perceived quality in the consumer's mind.

1. **Ghazi Zadeh and Hamayeli Mehrabani (2013)** studied green marketing strategies as a competitive advantage in the new era. Nowadays, companies are compelled to pay more attention to environmental issues due to environmental regulations, economic impacts, and increased public sensitivity to environmental matters. Many leading companies have been able to use this as a competitive advantage against their rivals by understanding the importance of environmental issues and the concerns of communities. This has been made possible by incorporating green marketing strategies into their planning. Therefore, in this article, we aim to pave the way for companies to use green marketing as a very important competitive advantage in the present era by theoretically examining issues related to green marketing, especially green marketing strategies.

2. **Ebad Askari and colleagues (2013)** investigated the impact of green products on the consumer purchasing decision process (a case study of consumers of disposable paper dishes in Torbat-e Jam city). Research shows that individuals have modified their purchasing behavior due to environmental issues. This study was conducted within the theoretical framework of consumer behavior theories and green marketing as a field study. The research aimed to examine the impact of green products on the consumer purchasing decision process. It is a survey in terms of method, descriptive in nature, and applied in terms of purpose. The statistical population of this research includes all consumers of disposable paper dishes in Torbat-e Jam city. The sampling method of the research is a random cluster based on geographical areas. After collecting data using a researcher-made questionnaire, descriptive and inferential statistics were used for data analysis. The research hypothesis was confirmed, showing that green products affect the consumer purchasing decision process. Therefore, suggestions for improving green marketing have been presented.

3. **Polonsky (2011)** addressed transformative green marketing: barriers and opportunities. Green marketing has failed to reach its potential in improving customers' living conditions but has improved natural ecosystems. The failure of this marketing is due to the inability of companies' customers and governments to have thinking systems that integrate macro marketing perspectives into very small decisions, thereby ignoring the superior view of humans in the natural environment. This article discusses why the three groups mentioned above had difficulties in accepting environmental issues, thus preventing the occurrence of real transformative green marketing. To examine these difficulties, three proposed activities should be implemented: (1) Marketers need to pay attention to new ways of creating connections that integrate environmental values and move away from financial scales that have no real environmental meaning. (2) Changing attitudes related to the environment highlights the importance of activity and inactivity, which should be based on increased education on the commonalities between humans and the environment. (3) Marketing needs to refocus its emphasis on satisfying desires by shifting away from acquiring goods, thereby enhancing how value is created in marketers. Implementing these changes will enable marketers to implement transformative green marketing practically so that both human conditions and the natural system in which humans exist are enhanced and aligned with transformative green marketing.

**Research Design**

**Research Methodology and Objective**: This study is applied in terms of its objective and descriptive survey in terms of its research method, falling under the category of correlational research.

**Sampling Method and Data Collection**: The study population includes managers and employees of export companies such as Kalleh, Solico, Pemina, Sorbon, Mac, Pol, and Casel, active during 2015-2016. According to the obtained statistics, their total number is 345, of which, based on Cochran's formula, approximately 182 individuals were selected for the study.

**Statistical Methods and Data Analysis Approach**: This research utilized the Green Marketing Strategy questionnaire, derived from Imam Gholi's (2014) research (including 7 questions on green products, 8 on green promotion, 3 on green distribution, 5 on green pricing), and Aaker's (1991) Brand Quality questionnaire (consisting of 8 questions). Since this research sampled from the population and the data were based on a 5-point Likert scale ranging from "Strongly Disagree" to "Strongly Agree," scoring for the questions was calculated from 1 to 5. After collection, the data from the questionnaires were transferred to SPSS 19 software for raw data analysis using descriptive statistics (frequency, percentage, tables) and inferential statistics (Pearson correlation).

**Reliability and Validity of the Research Questionnaires**: To assess the validity of the measurement tools used in this research, the opinions of university professors and experts were utilized. Since the items in this questionnaire were designed based on standard questionnaires, the questionnaire is considered to have satisfactory validity. In this research, a preliminary sample including 30 pre-test questionnaires was conducted to examine reliability. Then, using the data obtained from these questionnaires and with the help of SPSS software, the Cronbach's alpha reliability coefficient was calculated, which was 0.972 for the Green Marketing Strategy questionnaire, 0.945 for green product, 0.881 for green promotion, 0.806 for green distribution, 0.912 for green pricing, and 0.872 for the Brand Quality questionnaire. Therefore, the questionnaires have high internal reliability, correlation, and validity.

**Conceptual Model of Research and Hypotheses**: In this research, hypotheses are tested following the models of Hick and Yildan (2013) and Moskovi and colleagues (2015). The conceptual model of the current research is researcher-made and results from reviewing various models and literature (Figure 1).

Figure 1: Conceptual Model of Research

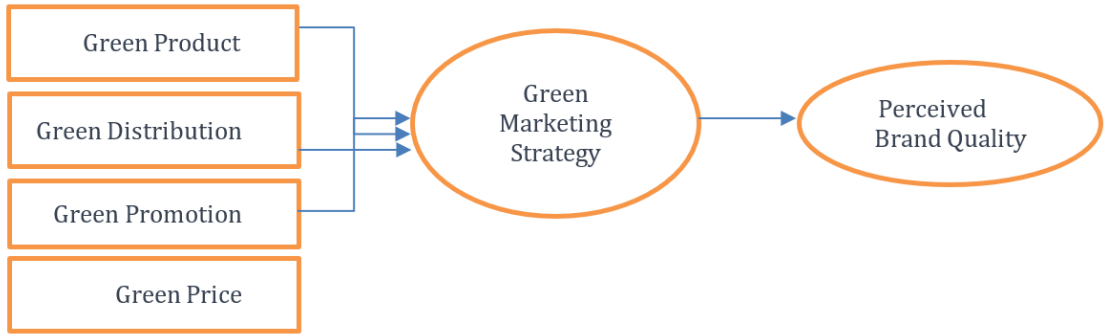

Main Hypothesis of the Research: Green marketing strategy has a significant effect on perceived brand quality.

Sub-Hypothesis 1: Green products have a significant effect on perceived brand quality.

Sub-Hypothesis 2: Green promotion has a significant effect on perceived brand quality.

Sub-Hypothesis 3: Green distribution has a significant effect on perceived brand quality.

Sub-Hypothesis 4: Green pricing has a significant effect on perceived brand quality.

**Findings**

The frequency and percentage distribution of the research sample group based on gender, educational qualification, and service history are presented in Table (1).

Table 1: Frequency and percentage distribution of the research sample group

| Variable | Subgroups | Frequency | Percentage Frequency |
|---|---|---|---|
| **Gender** | Male | 103 | 56.60% |
| | Female | 79 | 43.40% |
| **Age** | Less than 30 | 53 | 29.10% |
| | 31 to 40 | 61 | 33.50% |
| | 41 to 50 | 46 | 25.30% |
| | More than 50 | 22 | 12.10% |
| **Education** | Diploma and below | 31 | 17.00% |
| | Associate degree | 43 | 23.60% |
| | Bachelor's degree | 59 | 32.40% |
| | Graduate degree | 49 | 26.90% |
| **Work Experience** | Less than 1 year | 21 | 11.50% |

| | 1 to 5 years | 34 | 18.70% |
| | 6 to 10 years | 78 | 42.90% |
| | More than 10 years | 49 | 26.90% |

As indicated in Table 1, the majority of the study sample in terms of gender are men; in terms of educational level are those with a bachelor's degree; in terms of work experience are those with 6-10 years, and in terms of age are those between 31-40 years old.

Table 2: Pearson Correlation Test (Main Hypothesis)

| | | Green Marketing Strategy | Brand Quality |
|---|---|---|---|
| Green Marketing Strategy | Pearson Correlation Coefficient | 1 | .878 |
| | Significance Level | | .000 |
| | Sample Size | 182 | 182 |
| Brand Quality | Pearson Correlation Coefficient | .878 | 1 |
| | Significance Level | .000 | |
| | Sample Size | 182 | 182 |

The results from the Pearson Correlation Test presented in Table 2 indicate that the significance level is zero and less than 0.05. This demonstrates a significant relationship between green marketing strategy and brand quality, thereby confirming the main hypothesis of the research. Additionally, the Pearson correlation coefficient is 0.878, which signifies a positive and significant effect of the green marketing strategy on brand quality.

Table 3: Pearson Correlation Test (First Sub-Hypothesis)

| | | Brand Quality | Green Product |
|---|---|---|---|
| Brand Quality | Pearson Correlation Coefficient | 1 | .700 |
| | Significance Level | | .000 |
| | Sample Size | 182 | 182 |
| Green Product | Pearson Correlation Coefficient | .700 | 1 |
| | Significance Level | .000 | |
| | Sample Size | 182 | 182 |

The results from the Pearson Correlation Test in Table 3 indicate that the significance level is zero and less than 0.05. This demonstrates a significant relationship between green products and brand quality, thereby confirming the first sub-hypothesis of the research. Additionally, the Pearson correlation coefficient is 0.700, which signifies green products' positive and significant effect on brand quality.

Table 4: Pearson Correlation Test (Second Sub-Hypothesis)

| | | Brand Quality | Green Promotion |
|---|---|---|---|
| Brand Quality | Pearson Correlation Coefficient | 1 | .840 |
| | Significance Level | | .000 |
| | Sample Size | 182 | 182 |
| | Pearson Correlation Coefficient | .840 | 1 |
| Green Promotion | Significance Level | .000 | |
| | Sample Size | 182 | 182 |

The results from the Pearson Correlation Test in Table 4 indicate that the significance level is zero and less than 0.05. This demonstrates a significant relationship between green promotion and brand quality, thereby confirming the second sub-hypothesis of the research. The Pearson correlation coefficient is also 0.840, which signifies a positive and significant effect of green promotion on brand quality.

Table 5: Pearson Correlation Test (Third Sub-Hypothesis)

| | | Brand Quality | Green Distribution |
|---|---|---|---|
| Brand Quality | Pearson Correlation Coefficient | 1 | .920 |
| | Significance Level | | .000 |
| | Sample Size | 182 | 182 |
| Green Distribution | Pearson Correlation Coefficient | .920 | 1 |
| | Significance Level | .000 | |
| | Sample Size | 182 | 182 |

The results from the Pearson Correlation Test in Table 5 indicate that the significance level is zero and less than 0.05. This demonstrates a significant relationship between green distribution and brand quality, thereby confirming the third sub-hypothesis of the research. Additionally, the Pearson correlation coefficient is 0.920, which signifies a positive and significant effect of green distribution on brand quality.

Table 6: Pearson Correlation Test (Fourth Sub-Hypothesis)

| | | Brand Quality | Green Pricing |
|---|---|---|---|
| Brand Quality | Pearson Correlation Coefficient | 1 | .787 |
| | Significance Level | | .000 |
| | Sample Size | 182 | 182 |
| Green Pricing | Pearson Correlation Coefficient | .787 | 1 |
| | Significance Level | .000 | |
| | Sample Size | 182 | 182 |

The results from the Pearson Correlation Test in Table 6 indicate that the significance level is zero and less than 0.05. This demonstrates a significant relationship between green pricing and brand quality, confirming the research's fourth sub-hypothesis. Additionally, the Pearson correlation coefficient is 0.787, which signifies a positive and significant effect of green pricing on brand quality.

Table 7: Research Hypotheses Examination Results

| Hypothesis | Pearson Coefficient | Confirmation/Rejection | Type of Relationship |
|---|---|---|---|
| Main | 0.878 | Confirmed | Positive |
| Sub 1 | 0.700 | Confirmed | Positive |
| Sub 2 | 0.840 | Confirmed | Positive |
| Sub 3 | 0.920 | Confirmed | Positive |
| Sub 4 | 0.787 | Confirmed | Positive |

**Discussion and Conclusion**

The current research represents a comprehensive effort to assess the influence of green marketing strategies on brand quality within the food export sector in Tehran. The study meticulously examined various components of green marketing, including green products, green promotion, green distribution, and green pricing, and their collective impact on perceived brand quality. The empirical findings from this research are consistent with the hypotheses and corroborate the results of previous studies by Hick and Yildan (2013), Moghadam and Amirhosseini (2015), and Souri and colleagues (2015), reinforcing the notion that green marketing strategies are not only environmentally responsible but also beneficial for brand perception and quality.

The research underscores the importance of integrating environmental considerations into business strategies. It suggests that managers in the food export industry should prioritize maintaining and enhancing product quality while simultaneously focusing on environmental sustainability. This involves not only ensuring the high quality of products but also emphasizing the recycling and reintegration of unusable or returned products back into the production cycle. By doing so, companies can guarantee long-term product quality through robust guarantees and warranties. Moreover, this approach aids in preventing the entry of consumer waste into the environment, thereby contributing to ecological preservation.

Furthermore, the research recommends that companies should actively engage with their customers by allowing them periodic access to their production facilities. This transparency initiative would enable customers to directly observe and verify the quality of raw materials, the efficiency and modernity of production equipment and machinery, adherence to production processes, and the final output's compliance with environmental standards. Such initiatives not only enhance customer trust and satisfaction but also serve as a powerful marketing tool, showcasing the company's commitment to quality and environmental stewardship. This strategy is likely to attract a broader customer base, as modern consumers are increasingly inclined towards brands that demonstrate environmental responsibility and transparency.

In conclusion, the research highlights that green marketing strategies are not just a trend but a crucial aspect of modern business practices, especially in sectors like food exportation, where environmental impact and product quality are paramount. Companies that adopt and effectively implement these strategies are likely to see not only an improvement in brand perception and customer loyalty but also contribute positively to environmental sustainability, aligning their business goals with the broader objectives of social and ecological responsibility.